\documentclass[pra,twocolumn,floatfix,superscriptaddress]{revtex4}
\usepackage{graphicx}
\usepackage{amsthm}
\usepackage{amsmath}
\usepackage{amssymb}
\usepackage{bbm,amsfonts}
\usepackage{nomencl}
\usepackage{amstext} 
\usepackage{color}

\newcommand{\cE}{\mathcal{E}}

\newcommand{\Id}{\mathbbm{1}}

\newcommand{\braa}{\langle \psi |}
\newcommand{\kett}{|\psi\rangle}

\newcommand{\Hilb}{\mathcal{H}}

\newcommand{\Eop}{\mathcal{E}}

\newcommand{\tr}{{\mathrm{tr}}}

\newcommand{\cH}{\mathcal{H}}

\newcommand{\cD}{\mathcal{D}}
\newcommand{\cT}{\mathcal{T}}
\newcommand{\cA}{\mathcal{A}}
\newcommand{\cU}{\mathcal{U}}
\newcommand{\cF}{\mathcal{F}}
\newcommand{\cP}{\mathcal{P}}

\newcommand{\ket}[1]{\left| #1 \right\rangle}
\newcommand{\bra}[1]{\left\langle #1 \right|}
\newcommand{\braket}[2]{\left\langle #1 | #2 \right\rangle}
\newcommand{\braopket}[3]{\bra{#1}#2\ket{#3}}
\newcommand{\proj}[1]{| #1\rangle\!\langle #1 |}
\newcommand{\ketbra}[2]{| #1\rangle\!\langle #2 |}
\newcommand{\muFS}{\mu_{\scriptscriptstyle{F\!S}}}

\begin{document}

\title{Gate fidelity fluctuations and quantum process invariants}
\date{\today}
\author{Easwar Magesan}
\affiliation{Institute for Quantum Computing and Department of Applied Mathematics, University of Waterloo, Waterloo, Ontario N2L 3G1, Canada}	
\author{Robin Blume-Kohout}
\affiliation{Theoretical Division, Los Alamos National Laboratory, Los Alamos, NM, 87545, U.S.A.}
\author{Joseph Emerson}
\affiliation{Institute for Quantum Computing and Department of Applied Mathematics, University of Waterloo, Waterloo, Ontario N2L 3G1, Canada}	

\begin{abstract}
We characterize the quantum gate fidelity in a state-independent manner by giving an explicit expression for its variance. The method we provide can be extended to calculate all higher order moments of the gate fidelity. Using these results we obtain a simple expression for the variance of a single qubit system and deduce the asymptotic behavior for large-dimensional quantum systems. Applications of these results to quantum chaos and randomized benchmarking are discussed.

\end{abstract}


%

\maketitle

\section{Introduction}\label{Introduction}

The building blocks of a quantum computation are quantum logic \emph{gates}, unitary operations that are applied in a specific sequence to the physical systems that encode quantum information.  In theory, any quantum algorithm can be implemented with high precision by applying a correctly chosen sequence of gates -- but in practice, gates have errors.  In real experiments, we attempt to apply an ideal gate $\mathcal{U}$, but what really occurs is a noisy quantum operation $\cE$.  If we want to end up with a dynamical evolution close to the desired algorithm, $\cE$ had better be ``close'' to $\mathcal{U}$.

How close $\cE$ is to $\mathcal{U}$, operationally, depends on the state of the system they act on.  A state $\rho$ may evolve identically under $\cE$ and $\mathcal{U}$, or $\cE(\rho)$ and $\mathcal{U}(\rho)$ could be drastically different. The \textit{gate fidelity} is an experimentally relevant measure of how close $\cE$ and $\mathcal{U}$ are given the input state $\rho$. Often one wants to remove this state-dependence  because understanding quantum noise and designing error-resistant devices requires state-independent characterizations of the noise. One method for obtaining a state-independent quantity from the gate fidelity is to average it over all [pure] input states. This average gate fidelity, $\overline{\cF}$, provides a concise, useful measure of error.


However, the average provides no information about \emph{fluctuations} in the gate fidelity -- i.e., how the error varies over input states.  The magnitude of the fluctuations is a useful diagnostic.  It provides information about the worst-case fidelity, which is relevant for fault-tolerant design.  Large fluctuations also suggest that the average error may be dominated by a few very error-prone states, in which case addressing those states can produce dramatic improvements in average fidelity.  Large fluctuations may also indicate hidden high-fidelity information-preserving structures such as pointer bases or decoherence-free subspaces \cite{KNPV}.

In this paper, we calculate the variance of the gate fidelity analytically and discuss how it might be measured in experiments (as well as challenges to doing so!).  Moreover, we develop a general method for calculating higher moments of the gate fidelity, which can be applied to other purposes.  This problem has been considered before:  Ref. \cite{PMM} solved the special case where $\cE$ maps pure states into pure states.  Our calculation applies to general quantum operations.  We also apply it to two interesting specific cases: operations acting on a single qubit, and on very large systems.

We begin in Section \ref{sec:Background} by introducing background material and the framework we will use throughout the presentation. In Section \ref{sec:Ave}, we calculate the average gate fidelity as a warm-up, and confirm agreement with previous calculations.  We then calculate the variance in Section \ref{sec:Var}, and briefly discuss higher order moments in Section \ref{sec:Higher}. Next in Section \ref{sec:Single}, we calculate an explicit expression for gates on a single qubit as well as an explicit upper-bound that depends only on the dimension of the system which allows us to deduce the asymptotic behaviour of the variance. We conclude, in Section \ref{sec:Disc}, by discussing applications to randomized noise characterization, and to estimating fidelity decay under controlled perturbations of chaotic systems.

\section{Notation and Background}\label{sec:Background}

Before beginning, we set some notation for the rest of the paper. $\cH$ represents a system's Hilbert space, assumed in this paper to be of finite dimension $d$.  $L(\cH)$ is the set of all linear operators on $\cH$ (i.e., $d\times d$ matrices), and $\cD(\cH)$ is the set of (mixed) quantum states on $\cH$, containing all the positive, trace-1 operators in $L(\cH)$. Pure states are represented by elements of $\cH$ with $2$-norm equal to $1$, modulo phase factors (also known as complex projective space $\mathbb{CP}^{d-1}$).

\subsection{Quantum Operations and Representations of Linear Superoperators}\label{sec:Operations}

Quantum operations -- a.k.a. processes or channels -- describe the dynamical evolution of quantum systems over a period of time.  These dynamics may be reversible or irreversible, and they may even involve adding or discarding parts of the system, so the initial and final Hilbert spaces need not be identical.  Let $\cH_1$ and $\cH_2$ be Hilbert spaces of dimension $d_1$ and $d_2$, representing (respectively) the input and the output of a quantum dynamical process.
We will denote the set of \emph{all} linear (super)operators from $L(\Hilb_1)$ to $L(\Hilb_2)$ by $\mathcal{T}(\Hilb_1,\Hilb_2)$, and if $\cH_1=\cH=\cH_2$, denote it $\cT(\cH)$.

Not every linear superoperator $\cE$ is a valid quantum operation.  First, $\cE$ must preserve the trace of the input state $\rho_1\in\cD(\cH_1)$, for it represents total probability.  Furthermore, a superoperator $\cE$ that maps some positive semidefinite $\rho_1\geq0$ to a \emph{non}-positive operator is physically impossible -- it is not a \emph{positive} map.  In fact, to represent a valid operation, a superoperator must satisfy the even stronger condition of \emph{complete positivity}: given an ancillary system represented by a Hilbert space $\cA$, $\cE$ must map every joint state $\rho_{\cH_1,\cA}\in\cD(\cH_1\otimes\cA)$ to a positive semidefinite state of $\cH_2\otimes\cA$.  Complete positivity (CP), together with trace-preservingness (TP), is both necessary \emph{and} sufficient for $\cE$ to represent a valid quantum operation.

There are many ways to represent a CPTP map, and more generally a linear superoperator, which include the standard representation, Choi representation~\cite{Choi}, Kraus representation~\cite{Choi,Kraus}, $\chi$-representation~\cite{CN97} and Stinespring's representation~\cite{Stine}. A good reference for completely positive maps and their various representations is given by~\cite{Paul}. We briefly describe the standard, Choi, Kraus and $\chi$-representations for general linear superoperators as these representations will be used frequently throughout the rest of the presentation. We then show in the next subsection that the Choi and $\chi$-representations can be identified by choosing appropriate bases to write the respective representations in.

Let $\Eop \in \mathcal{T}(\Hilb_1,\Hilb_2)$ and $\{Q_i\}$, $\{R_j\}$ be bases for $L(\Hilb_1)$ and $L(\Hilb_2)$ respectively. The most straightforward representation of a linear superoperator comes from the observation that $L(\cH)$ is itself a Hilbert space of dimension $d^2$ under the Hilbert-Schmidt inner product,

\begin{equation}
\braket{A}{B} \equiv \tr \left(A^\dagger B\right).\nonumber
\end{equation}

\noindent The standard representation of $\Eop$ with respect to the above bases is the $d_2^2$ by $d_1^2$ matrix,

\begin{gather}
\Eop_{i,j}=\tr \left(R_i^{\dagger}\Eop(Q_j)\right).\nonumber
\end{gather}

\noindent While this representation is both natural and useful, complete positivity is not easily tested in this representation.  So we turn to the Choi representation.

The Choi representation for $\Eop$, denoted $C(\Eop)$, is the linear operator on $\Hilb_2 \otimes \Hilb_1$ given by,

\begin{equation}
C(\Eop)= \sum_{(a,b) \in \mathbb{Z}_{d_1}\times \mathbb{Z}_{d_1}}\Eop(|a\rangle\langle b|)\otimes |a\rangle \langle b|= \left(\Eop\otimes \mathcal{I}\right)(d_1\sigma)\label{eq:Choimatrixdef}
\end{equation}

\noindent where $\sigma$ is the maximally entangled Bell state,

\begin{gather}
\sigma = |\Psi\rangle \langle \Psi | = \left(\frac{1}{\sqrt{d_1}}\sum_{a=1}^{d_1}|a\rangle \otimes |a\rangle \right) \left(\frac{1}{\sqrt{d_1}}\sum_{b=1}^{d_1}\langle b | \otimes \langle b | \right).\nonumber
\end{gather}

 \noindent Note that the association $\Eop \rightarrow C(\Eop)$ is an isomorphism between $\mathcal{T}(\Hilb_1,\Hilb_2)$ and $L\left(\Hilb_2 \otimes \Hilb_1\right)$ and also that for any $\Eop_1$ and $\Eop_2$, $C(\Eop_1 \otimes \Eop_2)=C(\Eop_1)\otimes C(\Eop_2)$. From Eq. (\ref{eq:Choimatrixdef}), $\Eop$ is completely positive and trace-preserving if and only if $\frac{1}{d_1}C(\Eop)$ is a quantum state in $L(\Hilb_2 \otimes \Hilb_1)$. Hence, the mapping $\Eop \rightarrow \frac{1}{d_1}C(\Eop)$ is a linear isomorphism that is a bijection between quantum operations and quantum states. The state $J(\Eop):=\frac{1}{d_1}C(\Eop)$ is commonly called the \emph{Jamiolkowski state} associated to $\Eop$ and the isomorphism is known as the \emph{Choi-Jamiolkowski isomorphism}.

 When $C(\Eop)$ is written with respect to the basis $\{|a\rangle |b\rangle \langle c|\langle d|\}$ (where $a$ and $c$ range from $0$ to $d_2-1$, $b$ and $d$ range from $0$ to $d_1-1$, and we assume the right-most index varies fastest in tensor product state bases) the resulting matrix is called the \emph{Choi matrix}. The \emph{Jamiolkowski matrix} is naturally defined as the Choi matrix multiplied by $\frac{1}{d_1}$. Note that this definition does not imply the Choi matrix corresponds to simply block-constructing a matrix via $\left(\Eop\left(|i\rangle\langle j|\right)\right)_{i,j}$. This correspondence would hold however if we either assumed that the left-most index varies fastest in tensor product state bases or defined $C(\Eop) = \left(\mathcal{I} \otimes \Eop \right)(d_1\sigma)$.

A Kraus representation of the linear superoperator $\Eop$ can be obtained from $C(\Eop)$. By the singular value decomposition,

\begin{equation}
C(\Eop)=\sum_{i=1}^k |a_i\rangle\langle b_i| \nonumber
\end{equation}

\noindent where the $|a_i\rangle$ and $|b_i\rangle$ are proportional to the left and right singular vectors of $C(\Eop)$ respectively, and k is the rank of $C(\Eop)$. There is an obvious inner-product isomorphism between $L\left(\Hilb_1,\Hilb_2\right)$ with the Hilbert-Schmidt inner product and $\Hilb_2 \otimes \Hilb_1$ with the standard inner product, defined by $|a\rangle\langle b| \rightarrow \text{vec}\left(|a\rangle\langle b|\right) = |a\rangle \otimes |b\rangle$. If $A_i$ and $B_i$ are the unique linear operators in $L\left(\Hilb_1,\Hilb_2\right)$ satisfying $\text{vec}(A_i)=|a_i\rangle$ and $\text{vec}(B_i)=|b_i\rangle$ respectively, then for every $M \in L(\Hilb_1)$,

\begin{equation}
\Eop(M)=\sum_{i=1}^k A_iM B_i^{\dagger}.\nonumber
\end{equation}

\noindent The above expression is called a Kraus representation for $\Eop$ and, unlike the Choi representation, is not unique. If $\Eop$ is completely positive and trace preserving then $B_i=A_i$ for each i and $\sum_{i=1}^kA_i^{\dagger}A_i = \Id$.

Lastly, a useful representation in quantum process tomography is the $\chi$-representation of a quantum operation. If the linear superoperator $\Eop$ has Kraus operators $\{A_i, B_i\} \in L\left(\Hilb_1,\Hilb_2\right)$, and if $\{Q_j\}$ is a basis for $L\left(\Hilb_1,\Hilb_2\right)$, we can expand the Kraus operators in this basis and write the action of $\Eop$ on $M \in L(\Hilb_1)$ as,

\begin{eqnarray}
\Eop(M)&=&\sum_{i=1}^k A_iM B_i^{\dagger} \nonumber \\
&=&\sum_{i,j}\chi_{i,j}Q_iMQ_j^{\dagger}.\nonumber
\end{eqnarray}

\noindent The $d_1d_2$ by $d_1d_2$ matrix $\chi_{i,j}$ is called the $\chi$-matrix for $\Eop$ and is unique given the choice of basis $\{Q_i\}$ (it does not depend on the choice of Kraus operators for $\Eop$). We show next that $\chi_{i,j}$ written in the basis $\{Q_i\}$ can be identified with the Jamiolkowski representation written in a bipartite basis determined by the $\{Q_i\}$.


\subsection{Identifying the $\chi$ and Jamiolkowski Representations}

The $\chi$ and Jamiolkowski representations can be identified in the following manner: First, it is straightforward to show that if $\{Q_i\}$ is a basis for $L\left(\Hilb_1,\Hilb_2\right)$ then $\{\left(Q_i\otimes \Id\right) |\Psi\rangle\}$ is a basis for the bipartite space $\Hilb_2 \otimes \Hilb_1$. Next, if $\Eop \in \mathcal{T}(\Hilb_1,\Hilb_2)$ then,

\begin{eqnarray}
J(\Eop) &=& \frac{1}{d}\sum_{a,b} \Eop(|a\rangle\langle b|) \otimes |a\rangle\langle b| \nonumber \\
&=& \frac{1}{d}\sum_{a,b} \left(\sum_{i,j}\chi_{i,j}Q_i|a\rangle\langle b|Q_j^{\dagger} \right) \otimes |a\rangle\langle b| \nonumber \\
&=& \sum_{i,j}\chi_{i,j} \left(Q_i\otimes \Id\right) |\Psi\rangle\langle \Psi | \left(Q_j^{\dagger} \otimes \Id\right) \nonumber
\end{eqnarray}

\noindent and so the $\chi$-matrix of $\Eop$ relative to $\{Q_i\}$ is \emph{equal} to the Jamiolkowski state of $\Eop$ written with respect to the basis $\{\left(Q_i\otimes \Id\right) |\Psi\rangle\}$. Therefore there is no less of generality in writing $\chi$ to represent the linear operator $J(\Eop)$. Hence, throughout the rest of the presentation, ``$\chi_{i,j}$" will (unambiguously) refer to either $J(\Eop)$ written in the bipartite basis $\{\left(Q_i\otimes \Id\right) |\Psi\rangle\}$ or the $\chi$-matrix of $\Eop$ with respect to $\{Q_i\}$.



Note that for a quantum operation $\Eop$, even though $J(\Eop)$ as defined can be associated to a quantum state, writing $J(\Eop)$ with respect to $\{\left(Q_i\otimes \Id\right) |\Psi\rangle\}$
may produce a positive semidefinite matrix $\chi_{i,j}$ that does not have unit trace. It is straightforward to show however that if $\{Q_i\}$ is an orthogonal basis of $L(\Hilb)$ normalized
so that $\tr\left(Q_j^{\dagger}Q_i\right)=\delta_{i,j}d$, then $\chi_{i,j}$ itself is a positive semi-definite, trace-1 matrix.
A standard example of such a basis $\{Q_i\}$ is the set of (normalized) matrix units $\{\sqrt{d}\ketbra{k}{l}\}$, $k,l \in \mathbb{Z}_d$. In the next section we discuss other bases satisfying these conditions which will be more convenient for the calculations we deal with in this paper.

\subsection{Generalized bases and the partial trace/partial transpose} \label{sec:Genbases}

For qubits ($d=2$), the Pauli operators are an exceptionally convenient basis for $L(\cH)$.  The corresponding basis of bipartite states is the Bell basis.  In higher dimensions, it's generally not possible to pick a basis with \emph{all} the nice properties of the Pauli operators, but we can generalize most of them.

We will make extensive use of the existence of a Hermitian, orthogonal basis of matrices $\{P_a\}$ satisfying the following conditions:
\begin{eqnarray}\label{eq:basisconds}
\tr\left(P_aP_b\right) &=& d\delta_{a,b} \nonumber \\
P_a^\dagger &=& P_a \nonumber \\
P_0 &=& \Id.
\end{eqnarray}

\noindent In dimension $d=2$, the ubiquitous Pauli operators form such a basis; in higher dimensions, the generalized Gell-Mann operators~\cite{Geo} satisfy the conditions.  We require only the listed conditions -- most importantly, the fact that $P_0=\Id$ and therefore every other $P_k$ is traceless, which singles out the $\chi_{0,0}$ matrix element as a unitary invariant of $\cE$ -- so we will not specify a particular basis. The corresponding basis of bipartite states $\{\left(P_a\otimes \Id\right) |\Psi\rangle\}$ is orthonormal and because $P_0=\Id$ we have $\ket{\psi_0} = \ket\Psi$.  We will refer to this basis as the ``generalized Bell basis,'' though it does not by any means generalize \emph{all} the properties of the Bell states.

We will also make extensive use of the bipartite projector
$$\chi_0 = \proj{\psi_0} = \proj{\Psi}.$$
It is equal to $P_0 \otimes \Id \proj{\Psi}$ and is proportional to the Choi representation of the identity channel $\cE = \Id$, which motivates the notation $|\psi_0\rangle$ and $\chi_0$.  Moreover, it enables us to write expressions we derive using the basis $\{P_a\}$ in terms of quantities that are defined independently of any basis -- e.g., the unitary invariant $\chi_{0,0}$ mentioned above can be written as
$$\chi_{0,0} = \braopket{\psi_0}{\chi}{\psi_0} = \tr\left[\chi\chi_0\right].$$




Two important operations on bipartite matrices are the partial trace and partial transpose. For $A \in L(\Hilb \otimes \Hilb)$ these operations are denoted and defined as follows:

\begin{itemize}
\item The \emph{partial trace} over one subsystem: We will denote the partial trace over subsystem $i$ by $\tr_i[A]$.  When the partial trace is applied to a state, it generates the reduced density matrix of the remaining subsystem: $\rho_2 = \tr_1\rho$.
\item The \emph{partial transpose} over one subsystem:  For any bipartite matrix $A$, we will denote the partial transpose of $A$ with respect to the $i$th subsystem by $A^{T_i}$.  (Similarly, $A^T$ indicate the full transpose of $A$, with respect to both subsystems).  Partial transposition is \emph{not} a completely positive operation; in particular, it transforms many entangled states into negative matrices.  Interestingly, it appears naturally in our result.  Explicit expressions for the partial transpose operation on the subsystems are given by:
\begin{eqnarray}
A^{T_1} &=& \sum_{k,l=0}^{d-1} \left(\ketbra{k}{l} \otimes \Id\right) A \left(\ketbra{k}{l} \otimes \Id\right) \nonumber\\
A^{T_2} &=& \sum_{k,l=0}^{d-1} \left(\Id \otimes \ketbra{k}{l}\right) A \left(\Id \otimes \ketbra{k}{l}\right). \nonumber
\end{eqnarray}
\end{itemize}


\subsection{Gate Fidelity}\label{sec:Gatefidelity}

Suppose that $\cU$ is a unitary quantum operation (i.e., $\cU(\rho) = U\rho U^\dagger$ for some unitary $U$), and $\cE$ is a noisy implementation of $\cU$.  Then the \emph{gate fidelity} between $\cE$ and $\mathcal{U}$, given state $\rho$, is
\begin{equation}
\mathcal{F}_{\cE,\cU}\left(\rho\right)=\left(\tr\sqrt{\sqrt{\cE(\rho)}\cU(\rho)\sqrt{\cE(\rho)}}\right)^2. \nonumber
\end{equation}
It is simply the \emph{state fidelity} between $\cE\left(\rho\right)$ and $\cU\left(\rho\right)$, defined in general as
\begin{equation}
F\left(\rho,\sigma\right) = \left(\tr\sqrt{\sqrt{\rho}\sigma\sqrt{\rho}}\right)^2.\nonumber
\end{equation}
Fidelity measures indistinguishability:  $F=1$ means the states are identical, while $F=0$ implies that a single measurement can distinguish them perfectly.  Thus, the gate fidelity $\mathcal{F}_{\cE,\cU}(\rho)$ is a convenient measure of how distinguishable the actions of $\cU$ and $\cE$ are -- \emph{on the state $\rho$}.

If the input state is pure (so $\rho = \proj{\phi}$), then
\begin{equation}
\cF_{\cE,\cU}(\phi) = \tr\left[\mathcal{U}(\proj{\phi})\cE(\proj{\phi})\right].\nonumber
\end{equation}
If $\{K_j\}$ are Kraus operators for $\cE$, then we can rewrite this using the cyclic property of the trace as
\begin{eqnarray}
\mathcal{F}_{\cE,\mathcal{U}}(\phi) &=& \tr\left[U\proj{\phi} U^{\dagger} \sum_j K_j \proj{\phi} K_j^{\dagger}\right] \nonumber \\
&=& \tr\left[\proj{\phi} \: \mathcal{U}^{\dagger} \circ \cE (\proj{\phi})\right] \nonumber \\
&=& \tr\left[\proj{\phi} \Lambda (\proj{\phi})\right]\nonumber
\end{eqnarray}
where $\Lambda = \mathcal{U}^{\dagger} \circ \cE$ represents how much $\cE$ deviates from $\mathcal{U}$.

We would like a performance measure that removes the state dependence -- an invariant property of the gate fidelity's distribution.  If we focus our attention on pure input states, then this distribution is well defined, for the set of pure quantum states, $\mathbb{CP}^{d-1}$, admits a unique (and natural) invariant distribution.  It is the Fubini-Study (FS) measure~\cite{BZ} which we will denote by $\muFS$.  This is the Borel probability measure induced by the Fubini-Study metric on $\mathbb{CP}^{d-1}$ (ie. the unique unitarily invariant Haar probability measure on $\mathbb{CP}^{d-1}$).  The \emph{average} fidelity under $\muFS$ has been derived previously (see, e.g., Refs. \cite{EAZ, Nie02}), and we will derive it again in Section \ref{sec:Variance} as a warmup for computing higher moments of the gate fidelity distribution.

\subsection{Permutation Operators and the Symmetric Subspace}\label{sec:Perm}

To compute averages over a unitarily invariant measure we will begin by transforming polynomial functions of degree $k$ into linear functions on $k$ copies of the Hilbert space in question.  We will then rely on a simple and beautiful result called \emph{Schur-Weyl duality}, which states (in essence) that the actions of the unitary group and the permutation group (on such a $k$-fold tensor product) commute, and their irreducible representations (irreps) share a set of labels. Rather than discuss Schur-Weyl duality in detail, we will only introduce the tools that we need.  In this section, we will briefly discuss permutation operators, the symmetric group, the totally symmetric subspace of $\cH^{\otimes k}$, and a couple of technical results that will be useful later.

Let $\cH$ be a Hilbert space and $\cH^{\otimes k}$ a tensor product of $k$ copies of it.  If $S_k$ is the symmetric group on k objects and $\sigma \in S_k$ is a permutation, then there exists a unitary operator $\cP_{\sigma}$ that implements $\sigma$ on $\cH^{\otimes k}$:
\begin{equation*}
\cP_{\sigma}\left(\ket{\psi_1}\otimes ... \otimes \ket{\psi_k}\right) = \ket{\psi_{\sigma ^{-1}(1)}} \otimes ... \otimes \ket{\psi_{\sigma ^{-1}(k)}}.
\end{equation*}
The totally symmetric subspace of $\Hilb^{\otimes k}$ comprises all the states that are invariant under every such permutation operator -- or, to put it another way, it is the intersection of the $+1$ eigenspaces of all $\cP_{\sigma}$.  The projector onto this space is given by
\begin{equation}
\pi_{\text{sym}}(k,d) = \frac{1}{k!}\sum _{\sigma \in S_k}\cP_{\sigma}.\nonumber
\end{equation}

This projector appears in integrals over the unitary group, for the following reason (see Ref. \cite{RBSC}).  Suppose we take a state $\ket\psi\in\cH$, and then construct the projector onto its $k$-fold tensor product, $\proj\psi^{\otimes k}$.  This projector is a $+1$ eigenoperator of every permutation, so it lies in the totally symmetric subspace.  Now, if we take the \emph{average} of all such projectors according to the unitarily invariant measure $\muFS$ (denoted $\overline{\proj{\psi}^{\otimes k}}$), then we get an operator in $L\left(\cH^{\otimes k}\right)$ that: (i) is invariant under all unitaries $U^{\otimes k}$; (ii) is supported on the totally symmetric subspace; and (iii) has unit trace.  By Schur's Lemma, a unitarily invariant operator is a weighted sum of projectors onto irreducible representations of the unitary group.  The only such operators supported on the totally symmetric subspace are proportional to $\pi_{\text{sym}}$ itself.  Since $\overline{\proj{\psi}^{\otimes k}}$ has unit trace,
\begin{equation}\label{eq:symprojector}
\overline{\proj{\psi}^{\otimes k}} \equiv \int_{\psi\in CP^{d-1}}\proj\psi^{\otimes k}d\muFS = \frac{\pi_{\text{sym}}(k,d)}{\tr\left[\pi_{\text{sym}}(k,d)\right]}.
\end{equation}
The normalization constant is easy to evaluate by counting arguments.  The symmetric subspace of $\cH^{\otimes k}$ is spanned by the bosonic Fock states, $\ket{ n_1,n_2,\ldots, n_d}$, which are indexed by the number of particles $n_i$ in state $i$, subject to $\sum_i{n_i} = k$.  Counting such states, we get
\begin{eqnarray} \label{eq:symnorm}
\tr\left[\pi_{\text{sym}}(k,d)\right] &=& \binom{k+d-1}{d-1} \nonumber \\
&=& \frac{d(d+1)(d+2)\ldots(d+k-1)}{k!}.\label{eq:symnorm}
\end{eqnarray}

Suppose that we have $k$ operators $A_1,...,A_k$ in $L\left(\Hilb\right)$, and a permutation $\sigma \in S_k$ written as a product of disjoint cycles $\left(a_{1}...a_{r}\right)...\left(a_{q}...a_{k}\right)$. Then
\begin{equation}\label{eq:tracetensor}
\tr\left[\left(A_1\otimes...\otimes A_k\right) \: \cP_{\sigma}\right] = \tr\left[A_{a_1}...A_{a_r}\right]...\tr\left[A_{a_q}...A_{a_k}\right].
\end{equation}
So, to calculate $\tr\left[\left(A_1\otimes...\otimes A_k\right) \: \cP_{\sigma}\right]$, we can write $\sigma$ in cyclic notation, replace ``$i$'' with operator $A_i$, and replaces each $``(\: \:)"$ with $``\tr[\: \:]"$.

\section{Calculating the Variance of the Gate Fidelity}\label{sec:Variance}

We can use many of the tools described in the previous sections to calculate the variance of $\cF_{\cE,\cU}$ (hereafter denoted simply by $\cF$):
\begin{equation}
\text{Var}\left(\cF\right) = \overline{\cF^2} - \overline{\cF}^2.\nonumber
\end{equation}
As previously mentioned, the existence of a basis of linear operators $\{P_a\}$ with the properties listed in Eq. (\ref{eq:basisconds}) will play an important role in our derivation. Our ultimate goal is the expression (Eq. (\ref{eq:varchoi})), written in terms of the Choi representation $\chi$ for $\Lambda \equiv \cU^\dagger\circ\cE$.

\subsection{Average Gate Fidelity}\label{sec:Ave}

To determine $\text{Var}(\cF)$, we need to calculate both $\overline{\cF}$ and $\overline{\cF^2}$.  Fortunately, the tools in the previous section can be used calculate \emph{any} moment of $\cF$, although the calculation of $\overline{\cF^n}$ gets rapidly harder with increasing $n$.  So we begin with $\overline{\cF}$, which is already well-known \cite{Nie02}, as a sort of warmup.

We begin by expanding the state-dependent gate fidelity in terms of $\Lambda$'s Kraus operators $\{K_i\}$,
\begin{eqnarray}
\mathcal{F}(\proj\psi) &=& \tr\left[\Lambda(\proj\psi)\cdot\proj\psi \right] \nonumber \\ \noalign{\vskip 2mm}
&=& \sum_i \tr\left[K_i\proj\psi\right] \tr\left[K_i^{\dagger}\proj{\psi}\right] \nonumber\\
&=& \sum_i \tr\left[\left(K_i \otimes K_i^{\dagger}\right) \proj\psi\otimes\proj\psi \right] \nonumber.
\end{eqnarray}
This expression is a Hilbert-Schmidt inner product between (i) a term including all the Kraus operators, and (ii) a term including all the $\ket\psi$-dependence.  To average over $\psi$, we need only average the second term, using Eq. (\ref{eq:symprojector}):
\begin{eqnarray}
\overline{\mathcal{F}} &=& \int{\mathcal{F}(\proj\psi) d\psi} \nonumber \\
&=& \sum_i \tr\left[\left(K_i \otimes K_i^{\dagger}\right) \overline{\proj{\psi} \otimes \proj{\psi}}\right] \nonumber \\
&=& \sum_i \tr\left[\left(K_i \otimes K_i^{\dagger}\right)\frac{\pi_{\text{sym}}(2,d)}{\tr\left(\pi_{\text{sym}}(2,d)\right)}\right].\nonumber
\end{eqnarray}
We now expand $\pi_{\text{sym}}(2,d)$ as a sum of permutation operators, invoke Eq. (\ref{eq:tracetensor}) to evaluate the traces, and use Eq. (\ref{eq:symnorm}) to evaluate the normalization:
\begin{eqnarray}
\overline{\mathcal{F}} &=& \frac{1}{2\tr\left[\pi_{\text{sym}}(2,d)\right]}\sum_i\sum _{\sigma \in S_2}\tr\left[\left(K_i \otimes K_i^{\dagger}\right)\cP_{\sigma}\right] \nonumber \\
&=& \frac{\sum _i\left(\tr\left[K_i\right]\tr\left[K_i^{\dagger}\right]\right) + d}{d^2+d}.\nonumber
\end{eqnarray}
We can also write $\overline{\cF}$ in terms of the Choi representation $\chi$ of $\Lambda$. Since $\Lambda(\rho)= \sum _{l,m} \chi _{l,m} P_l \rho P_m$, the same calculation yields
\begin{eqnarray}
\overline{\mathcal{F}} &=& \frac{2}{d^2 + d} \sum_{l,m} \tr\left[\chi_{l,m} \left(P_l \otimes P_m\right) \pi_{\text{sym}}(2,d)\right] \nonumber \\
&=& \frac{1}{d^2 + d}\sum _{l,m} \chi _{l,m}\left(\tr\left[P_l\right]\tr\left[P_m\right] + \tr\left[P_lP_m\right]\right)\nonumber\\
&=& \frac{\chi _{0,0}d + 1}{d+1},\nonumber
\end{eqnarray}
which agrees with the results from Refs.~\cite{EAZ,Nie02}.  Recalling that $\chi_{0,0} = \tr\left[\chi\chi_0\right]$ we have,
\begin{equation}\label{eq:Fchi}
\overline{\mathcal{F}} = \frac{\tr\left[\chi\chi_0\right]d+1}{d+1}.
\end{equation}
We observe that $\tr\left[\chi\chi_0\right]$ represents the overlap of $\Lambda$ with the identity channel, and therefore how much $\Lambda$ leaves the input state unchanged.  It is also a unitary invariant of $\Lambda$; $\chi_{0,0}$ does not change if we rotate $\Lambda$ by a unitary channel $\cU$, mapping $\Lambda \to \cU^{-1}\circ\Lambda\circ\cU$.

\subsection{Variance of the Fidelity}\label{sec:Var}

Now, let's tackle the calculation of $\overline{\cF^2}$.  As done previously, we expand $\cF^2$ in terms of $\Lambda$'s Kraus operators,
\begin{gather}
\mathcal{F}^2(\proj\psi) = \tr\left[\Lambda(\proj\psi)\cdot\proj\psi\right]^2 \nonumber\\ \noalign{\vskip 2mm}
= \sum_i \tr[K_i\proj{\psi}]\tr[K_i^{\dagger}\proj{\psi}] \sum_j \tr[K_j\proj{\psi}]\tr[K_j^{\dagger}\proj{\psi}] \nonumber \\
= \sum_{i,j} \tr\left[\left(K_i \otimes K_i^{\dagger} \otimes K_j \otimes K_j^{\dagger}\right)\cdot\proj{\psi}^{\otimes 4}\right]\nonumber,
\end{gather}
and then use Eq. (\ref{eq:symprojector}) to simplify the average, $\overline{\cF^2}$, as

\begin{gather}
\overline{\mathcal{F}^2} = \sum_{i,j} \tr\left[\left(K_i \otimes K_i \otimes K_j \otimes K_j\right) \overline{\proj{\psi}^{\otimes 4}} \right] \nonumber \\
= \sum_{i,j} \tr\left[\left(K_i \otimes K_i^{\dagger} \otimes K_j \otimes K_j^{\dagger}\right) \frac{\pi_{\text{sym}}(4,d)}{\tr\left[\pi_{\text{sym}}(4,d)\right]}\right].\nonumber
\end{gather}
Finally, we write $\pi_{\text{sym}}(2,d)$ as a sum of permutation operators

\begin{equation}\label{eq:Fsquaredkraus}
\overline{\mathcal{F}^2} = \frac{\sum_{i,j}\sum _{\sigma \in S_4} \tr\left[\left(K_i \otimes K_i^{\dagger} \otimes K_j \otimes K_j^{\dagger}\right)\cP_{\sigma}\right]}{d(d+1)(d+2)(d+3)}, \end{equation}

\noindent invoke Eq. (\ref{eq:tracetensor}) to evaluate the traces, and use Eq. (\ref{eq:symnorm}) to evaluate the normalization:

\begin{equation*}
\overline{\mathcal{F}^2} = \frac{
\left(\begin{array}{l}
\ \ \sum_{i,j}\Big(\tr\left[K_i\right]\tr\left[K_i^{\dagger}\right]\tr\left[K_j\right]\tr\left[K_j^{\dagger}\right] \vspace{0.05in} \\
\: \: \: \: \: \: \: \: \: \: \: \: \: \: \: \: \: \: + \: \:  \tr\left[K_i K_j^{\dagger}\right]\tr\left[K_i^{\dagger}\right]\tr\left[K_j\right] + ...\Big)
\end{array}\right)
}{d(d+1)(d+2)(d+3)}.
\end{equation*}


\noindent There are 24 products of traces in the sum, each corresponding to one of the 4! permutations of 4 objects, so the ellipsis in the last equation represents 22 more terms.

In this case, it's more productive to use the basis $\{P_i\}$ and write $\cF$ using the $\chi$ matrix.  The same calculation then yields
\begin{equation}\label{eq:Fsquaredtotal}
\overline{\mathcal{F}^2} = \frac{
\left(\begin{array}{l}
\ \ \sum_{l,m,n,r} \chi_{l,m} \chi _{n,r}\Big(\tr\left[P_l\right]\tr\left[P_m\right]\tr\left[P_n\right]\tr\left[P_r\right] \vspace{0.05in} \\
\: \: \: \: \: \: \: \: \: \: \: \: \: \: \: \: \: \: + \: \: \tr\left[P_lP_r\right]\tr\left[P_m\right]\tr\left[P_n\right] + ...\Big)
\end{array}\right)
}{d(d+1)(d+2)(d+3)}.
\end{equation}


\noindent By writing out all 24 terms in the summation (excluded here for reasons of space and extreme tediousness), we can use the assumed properties of the basis $\{P_i\}$ to simplify this expression to
\begin{equation}
\overline{\mathcal{F}^2} =  \frac{
\left(\begin{array}{l}
\ \ d^4\tr[\chi\chi_0]^2 \vspace{0.05in}\\
  + \: d^3\tr\left[\chi_0\left(2\chi^2 + \chi\chi^T + \chi^T\chi + 2\chi\right)\right]\vspace{0.05in} \\
  + \: d^2\left(4\tr[\chi\chi_0] + \tr\left[\chi\chi^T\right] + \tr\left[\chi^2\right] + 1\right)\vspace{0.05in}\\
  + \: d\left(2\sum_l\tr\left[\left(\chi_{l,0}+\chi_{0,l}\right)P_l\Lambda\left(\Id\right)\right]+3\right)\vspace{0.05in}\\
  + \: 2\tr\left[\sum_{l,m}\chi_{l,m}P_l\Lambda(P_m)\right]  + \tr\left[\left(\Lambda\left(\Id\right)\right)^2\right]
\end{array}\right)
}{d(d+1)(d+2)(d+3)}.\label{eq:Fsquaredchi1}
\end{equation}

All but three of the terms in Eq. (\ref{eq:Fsquaredchi1}) are expressed solely in terms of $\Lambda$'s $\chi$-matrix.  The exceptions are:
\begin{itemize}
\item $\tr\left[\left(\Lambda\left(\Id\right)\right)^2\right]$ which comes from terms of the form $\sum_{lmnr}\chi_{l,m} \chi_{n,r}\tr[P_lP_mP_nP_r]$ that are produced by 4-cycle permutations like $\sigma = (1234)$.
\item $2d\tr\left[\sum_l\left(\chi_{l,0}+\chi_{0,l}\right)P_l\Lambda\left(\Id\right)\right]$ which comes from terms of the form $\sum_{lmnr}\chi_{l,m}\chi_{n,r}\tr[P_l P_n P_r]\tr[P_m]$ that are produced by 3-cycle permutations like $\sigma = (123)(4)$.
\item $2\tr\left[\sum_{l,m}\chi_{l,m}P_l\Lambda(P_m)\right]$ which comes from terms of the form $\sum_{lmnr}\chi_{l,m} \chi_{n,r}\tr[P_lP_nP_mP_r]$ that are produced by 4-cycle permutations like $\sigma = (1324)$.
\end{itemize}
Our next order of business is to rewrite these quantities in terms of the Choi representation $\chi$ using $\chi_0$, the partial transpose, and the partial trace (see Section \ref{sec:Background}).

The first term is easy. It's straightforward to verify that
$$\Lambda\left(\frac{\Id}{d}\right) = \tr_2\chi,$$
so
\begin{equation}
\tr\left[\left(\Lambda\left(\Id\right)\right)^2\right] = d^2\tr\left[(\tr_2\chi)^2\right]. \label{eq:FirstExceptionalTerm}
\end{equation}

We can rewrite the second term using the \\
non-Hermitian operator
\begin{eqnarray*}
\chi\chi_0 &=& \sum_{l,m}\chi_{l,m}\left(P_l\otimes \Id\right)\chi_0\left(P_m \otimes \Id\right)\chi_0 \nonumber \\
&=& \sum_l \chi_{l,0} \left(P_l \otimes \Id\right) \chi_0 \\
&=& \frac{1}{d}\sum_{l,i,j} \chi_{l,0} P_l \ketbra{i}{j} \otimes \ketbra{i}{j}
\end{eqnarray*}
and its adjoint $\chi_0\chi$.  Partial tracing over the second (ancillary) system yields
\begin{eqnarray*}
\tr_2(\chi\chi_0) &=& \frac{1}{d} \sum_l\chi_{l,0}P_l, \\
\tr_2(\chi_0\chi) &=& \frac{1}{d} \sum_l\chi_{0,l}P_l,
\end{eqnarray*}
which provides the following expression for the second term:
\begin{eqnarray}
\tr \left[\left(\sum_{l}\left(\chi_{l,0}+\chi_{0,l}\right)P_l\right)\Lambda \left(\Id\right)\right] = \nonumber \\
d^2\tr \left[\tr_2\left(\chi\chi_0 + \chi_0\chi\right) \tr_2\chi\right].\label{eq:SecondExceptionalTerm}
\end{eqnarray}

To rewrite the third exceptional term, we apply a few more tricks.  First, we observe that for any bipartite operator $A\otimes B$,
\begin{equation}
\tr\left[\chi^{T_2} (A\otimes B)\right] = \frac{1}{d}\tr\left[A \Lambda(B)\right].\nonumber
\end{equation}
This is easily shown from the definition of $\chi$.  Next, we note that since $\chi_0 = \frac{1}{d}\sum_{l,m}{\chi_{lm}\ketbra{l}{m}\otimes\ketbra{l}{m}}$, its partial transpose (over either subsystem) is
\begin{equation}
\chi_0^{T_1} = \chi_0^{T_2} = \frac{1}{d}\sum_{l,m}{\ketbra{l}{m}\otimes\ketbra{m}{l}}.\nonumber
\end{equation}
This bipartite operator is proportional to the unitary SWAP gate (which we denote $S$), which maps $\ket{l}\otimes\ket{m}\to\ket{m}\otimes\ket{l}$.  Now, consider the operator $S(S\chi)^{T_1}$, which can be written out as:
\begin{eqnarray*}
S(S\chi)^{T_1} &=& \frac{1}{d}S\left(S\sum_{ijlm}{\chi_{lm} P_l\ketbra{i}{j}P_m \otimes\ketbra{i}{j}}\right)^{T_1} \\
&=& \frac{1}{d}S\left(\sum_{ijlm}{\chi_{lm} \ketbra{i}{j}P_m \otimes P_l\ketbra{i}{j}}\right)^{T_1} \\
&=& \frac{1}{d}S\left(\sum_{ijlm}{\chi_{lm} P_m^T\ketbra{j}{i} \otimes P_l\ketbra{i}{j}}\right) \\
&=& \frac{1}{d}\sum_{ijlm}{\chi_{lm} P_l\ketbra{i}{i} \otimes P_m^T\ketbra{j}{j}} \\
&=& \frac{1}{d}\sum_{lm}{\chi_{lm} P_l\otimes P_m^T}.
\end{eqnarray*}
Together, these two observations imply that
\begin{equation}\label{eq:thirdrelate}
\tr\left[ \chi^{T_2} \left(S(S\chi)^{T_1}\right)^{T_2}\right] = \frac{1}{d^2}\sum_{lm}{\chi_{lm}\tr[P_l\Lambda(P_m)]},
\end{equation}
but $\tr\left[X^{T_2}Y^{T_2}\right] = \tr\left[XY\right]$ (just as for the full transpose), so the two partial transposes cancel.  Substituting in $S = d\chi_0^{T_1}$, we get the following expression for the third term:

\begin{eqnarray} \label{eq:chiandS}
\sum_{l,m}{\chi_{l,m}\tr\left[P_l\Lambda(P_m)\right]} &=& d^4\tr\left[ \chi \chi_0^{T_1} (\chi_0^{T_1}\chi)^{T_1}\right] \nonumber \\ \noalign{\vskip 2mm}
&=& d^4 \tr \left[\left(\chi_0^{T_1}\chi \right)^{\dagger} \left(\chi_0^{T_1}\chi \right)^{T_1}\right] \nonumber \\ \noalign{\vskip 2mm}
&=& d^4 \tr \left[\left(\chi \chi_0^{T_2}\right)^{\dagger} \left(\chi \chi_0^{T_2}\right)^{T_2}\right].
\end{eqnarray}

Hence if,

\begin{eqnarray}
a_1&=&\tr \left(\chi\chi_0\right)^2 + 2\tr \left[\left(\chi \chi_0^{T_2}\right)^{\dagger} \left(\chi \chi_0^{T_2}\right)^{T_2}\right], \nonumber \\ \noalign{\vskip 2mm}
b_1&=&2\tr \left(\chi^2\chi_0\right) + \tr \left(\chi\chi^T\chi_0\right) + \tr \left(\chi^T\chi\chi_0\right) + 2\tr \left(\chi\chi_0\right) \nonumber \\ \noalign{\vskip 2mm}
& \: & + 2\tr \left[\tr_2\left(\chi\chi_0 + \chi_0 \chi\right)\tr_2\left(\chi\right)\right], \nonumber \\ \noalign{\vskip 2mm}
c_1&=&4\tr \left(\chi\chi_0\right) + \tr \left(\chi\chi^T\right) + \tr \left(\chi^2\right) + 1 + \tr \left[\left(\tr_2\chi\right)^2\right], \nonumber \\ \noalign{\vskip 2mm}
d_1&=&3.\nonumber
\end{eqnarray}

\noindent then,

\begin{gather}
\overline{\mathcal{F}^2} =  \frac{a_1d^4+b_1d^3+c_1d^2+d_1d}{d^4 + 6d^3 + 11d^2 + 6d}.\label{eq:Fsquaredchoi}
\end{gather}

%

\noindent From Eq. (\ref{eq:Fchi}) we have,

\begin{equation}\label{eq:Fbarsquaredchoi}
\overline{\mathcal{F}}^2 = \frac{a d^2 + b d + 1}{d^2 + 2d + 1}
\end{equation}

\noindent where,

\begin{eqnarray}
a&=&\tr\left(\chi\chi_0\right)^2, \nonumber \\ \noalign{\vskip 2mm}
b&=&2 \tr\left(\chi\chi_0\right).\nonumber
\end{eqnarray}

\noindent Taken together, Eq.'s (\ref{eq:Fsquaredchoi}) and (\ref{eq:Fbarsquaredchoi}) give the following expression for $\text{Var}\left(\mathcal{F}\right)$,

\begin{gather}\label{eq:varchoi}
 \text{Var}\left(\mathcal{F}\right) =  \frac{a_2d^5+b_2d^4+c_2d^3+d_2d^2 + e_2d + f_2}{(d+1)^3(d+2)(d+3)}
\end{gather}

\noindent where,

\begin{eqnarray*}
a_2&=&a_1-a, \nonumber \\
b_2&=&b_1+2a_1-b-6a, \nonumber \\
c_2&=&a_1+2b_1+c_1 -11a - 6b-1, \nonumber \\
d_2&=&b_1 + 2c_1 + d_1 -6a -11b -6 \nonumber \\
e_2&=&c_1 + 2d_1 - 11 - 6b \nonumber \\
f_2&=&d_1-6.
\end{eqnarray*}

\section{Higher Order Moments}\label{sec:Higher}

We briefly discuss how to calculate both the higher order moments $\overline{\mathcal{F}^m}$ and central moments $\overline{\left(\mathcal{F}-\overline{F}\right)^m}$ of the gate fidelity $\mathcal{F}$. We have already given a detailed analysis of the $m=1$ and $m=2$ cases, and have provided explicit expressions for $\overline{\mathcal{F}}$, $\overline{\mathcal{F}^2}$, and $\text{Var}(\mathcal{F})=\overline{\mathcal{F}^2}-\overline{\mathcal{F}}^2$ in terms of the Jamiolkowski state of a quantum operation (note that the first central moment is just $\overline{\mathcal{F}}$). The central moments contain valuable information about the distribution of the gate fidelity. The second central moment (variance) is a measure of the spread of the distribution, the third central moment measures the skewness, and so on. Since the m'th central moment is just $\overline{\left(\mathcal{F}-\overline{F}\right)^m}$ and we have an expression for $\overline{\mathcal{F}}$, the expression for the m'th central moment is easily obtained if each of $\overline{\mathcal{F}^k}$ is known for $k=1,...,m$.

For $m \in \mathbb{N}$, the m'th power of $\mathcal{F}$, $\mathcal{F}^m$, has action on pure state $\proj{\psi}$,

\begin{gather*}
\mathcal{F}^m\left(|\psi\rangle\langle \psi|\right) = \tr\left[\Lambda\left(\kett\braa\right)\kett\braa\right]^m \nonumber \\ \noalign{\vskip 2mm}
= \sum_{i_1}\tr\left[\left(K_{i_1}\otimes K_{i_1}^{\dagger}\right)\kett \braa\otimes \kett \braa\right]...\nonumber \\ \noalign{\vskip 2mm}
 \: \: \: \: \:  \sum_{i_m}\tr\left[\left(K_{i_m}\otimes K_{i_m}^{\dagger}\right)\kett \braa\otimes \kett \braa\right]\nonumber \\ \noalign{\vskip 2mm}
= \sum_{i_1,...,i_m}\tr\left[\left(K_{i_1}\otimes K_{i_1}^{\dagger}\otimes ... \otimes K_{i_m}\otimes K_{i_m}^{\dagger}\right)\kett\braa^{\otimes m}\right].\nonumber
\end{gather*}

\noindent In an analogous method to that used in calculating an expression for the variance we have $\overline{\mathcal{F}^m}$ is given by,

\begin{equation*}
\frac{\displaystyle\sum\limits_{i_1,...,i_m}\tr\left[\left(K_{i_1}\otimes K_{i_1}^{\dagger}\otimes ... \otimes K_{i_m}\otimes K_{i_m}^{\dagger}\right)\pi_{sym}\left(2m,d\right)\right]}{\tr\left[\pi_{sym}\left(2m,d\right)\right]},\nonumber \\
\end{equation*}

\noindent and using the results regarding permutation operators and the symmetric subspace described in Sec. \ref{sec:Perm} we obtain,

\begin{equation*}
\overline{\mathcal{F}^m} = \frac{
\left(\begin{array}{l}
\ \ \displaystyle\sum\limits_{i_1,...,i_m}\Bigg\{\tr\left(K_{i_1}\right)\tr\left(K_{i_1}^{\dagger}\right)...\tr\left(K_{i_m}\right)\tr\left(K_{i_m}^{\dagger}\right)  \\
\: \: \: \: \: \: \: \: \: \: \: \: \: \: \: \: \: \: \: \: \: +...+ \tr\left(K_{i_1}K_{i_1}^{\dagger}...K_{i_m}K_{i_m}^{\dagger}\right)\Bigg\}
\end{array}\right)
}
{\left(2m\right)!{2m+d-1 \choose d-1}}
\end{equation*}


\noindent where again the $\{K_i\}$ are a set of Kraus operators for $\Lambda$. There are $(2m)!$ terms in the sum corresponding to the fact that there are $(2m)!$ elements in the symmetric group $S_{2m}$ and we have used the fact that,

\begin{equation}
\tr\left[\pi_{sym}\left(2m,D\right)\right] = {2m+d-1 \choose d-1}.\nonumber
\end{equation}

\noindent Expanding the $K_i$ in terms of the basis $\{P_i\}$ with the previously discussed properties gives,

\begin{equation*}
\overline{\mathcal{F}^m} = \frac{
\left(\begin{array}{l}
\ \ \displaystyle\sum \limits_{i_{1_1},i_{1_2},..,i_{m_1},i_{m_2}}\displaystyle\prod\limits_{j=1}^m\chi_{i_{j_1},i_{j_2}}\Bigg\{\tr\left(P_{i_{1_1}}\right)\tr\left(P_{i_{1_2}}\right)...\\
\: \: \: \: \: \: \: \: \: \: \: \: \: \: \: \: \: \: \: \: \: \: \: \: \: \: \: \: \: \: \: \: \: \: \: \tr\left(P_{i_{m_1}}\right)\tr\left(P_{i_{m_2}}\right)+...\Bigg\}
\end{array}\right)
}
{(2m)!{2m+d-1 \choose d-1}}
\end{equation*}

\noindent which can be written in terms of $\chi$,

\begin{equation*}
\overline{\mathcal{F}^m}= \frac{\left(\tr\left(\chi \chi_0\right)^m d^{2m}+...\right)}{(2m)! {2m+d-1 \choose d-1}}.
\end{equation*}

\section{The Single Qubit and Upper Bounds on the Variance}\label{sec:Single}

In this section we analyze the behavior of $\text{Var}\left(\mathcal{F}\right)$ in two useful cases, that of a single qubit (d=2) and as d grows to $\infty$. The calculations in both cases are straightforward but tedious and so are contained in the appendix. We first look at the case of a single qubit.

For a qubit system, one can obtain much simpler equations for $\text{Var}\left(\mathcal{F}\right)$ than Eq. (\ref{eq:varchoi}) (see Sec. \ref{sec:Varsingle}). The calculation involves starting from Eq. (\ref{eq:Fsquaredtotal}), grouping certain terms together, and considering various cases. The result of the calculation is that the second moment of $\mathcal{F}$ is given by,


\begin{equation}
\overline{\mathcal{F}^2} = \frac{
\left(\begin{array}{l} \vspace{0.05in}
 -48\tr \left(\chi\chi_0\right)^2 + 64\tr \left(\chi\chi_0\right) \\ \vspace{0.05in}
+ \: 24\tr \left(\chi\chi^T\chi_0 + \chi^T\chi\chi_0\right) + 32\tr\left(\chi^2\chi_0\right) \\ \vspace{0.05in}
+ \: 4\tr\left(\chi\chi^T\right)+12\tr\left(\chi^2\right) + 4\tr \left[\left(\tr_2\chi\right)^2\right] + 6\nonumber
\end{array}\right)
}{120}.
\end{equation}


\noindent Using Eq. (\ref{eq:Fbarsquaredchoi}) we obtain the following particularly simple analogue of Eq. (\ref{eq:varchoi}),

\begin{eqnarray}\label{eq:varsinglequbit}
\text{Var}\left(\mathcal{F}\right)
&=& - \frac{11}{180} + \frac{4}{45}\tr \left(\chi\chi_0\right) - \frac{38}{45}\tr \left(\chi\chi_0\right)^2 \nonumber \\
&\:& + \frac{4}{15}\tr\left(\chi^2\chi_0\right) + \frac{1}{10}\tr \left(\chi^2\right) \nonumber \\
&\:& + \frac{1}{5}\tr \left(\chi\chi^T\chi_0 + \chi^T\chi\chi_0\right)\nonumber \\
&\:&  +\frac{1}{30}\left(\tr \left(\chi\chi^T\right) + \tr \left[\left(\tr_2\chi\right)^2\right]\right).
\end{eqnarray}

\noindent Note that if $\Lambda = \mathcal{U}^{\dagger}\circ \cE$ is a Pauli channel then $\chi$ is diagonal and the variance takes the form,

\begin{gather}
\text{Var}\left(\mathcal{F}\right)
= - \frac{2}{45} + \frac{4}{45}\tr \left(\chi\chi_0\right) - \frac{8}{45} \tr \left(\chi\chi_0\right)^2 + \frac{2}{15}\tr\left(\chi^2\right).\nonumber
\end{gather}

It is relatively straightforward to obtain a generic upper-bound on $\text{Var}\left(\mathcal{F}\right)$ that holds for any $d$ and allows us to deduce the behavior of $\text{Var}\left(\mathcal{F}\right)$ in large dimensions (see Sec. \ref{sec:Large}). The idea is to use a suitable expression for the variance and bound the coefficients of the powers of d. The result is that,

\begin{equation}\label{eq:varupperbound}
\text{Var}(\cF) \leq \frac{4d^3 + 4d^{\frac{5}{2}} + 9d^2 + 4d^{\frac{3}{2}} + 5d}{\left(d+1\right)^2\left(d+2\right)\left(d+3\right)}.
\end{equation}

\noindent As a simple corollary, comparing powers of $d$ in the numerator and denominator of Eq. (\ref{eq:varupperbound}), we see that for large $d$,

\begin{eqnarray}
\text{Var}\left(\mathcal{F}\right) &\sim& O\left(\frac{1}{d}\right). \label{eq:asymptotbehav}
\end{eqnarray}

\noindent We again emphasize that Eq. (\ref{eq:varupperbound}) is completely general: for \emph{any} quantum operation $\cE$ and \emph{any} unitary operation $\cU$ acting on $L\left(\mathbb{C}^d\right)$, the gate fidelity between $\cE$ and $\cU$ has variance that satisfies Eq. (\ref{eq:varupperbound}). 



\section{Discussion}\label{sec:Disc}

We have given a method for calculating all moments of the gate fidelity $\mathcal{F}_{\cE,\mathcal{U}}$ between a unitary $\mathcal{U}$ and a quantum operation $\cE$. Using this method we have obtained a closed form expression for $\text{Var}\left(\mathcal{F}_{\cE,\mathcal{U}}\right)$ in terms of the Choi representation for $\Lambda = \mathcal{U}^{\dagger}\circ \cE$. A simple expression for the variance is given in the single qubit case and an explicit upper-bound for the variance is given for all $d$. This upper-bound shows that for large quantum systems the variance scales as $O\left(\frac{1}{d}\right)$ for any $\cE$ and $\cU$.

There is growing interest in estimating partial information about the unknown noise affecting the implementation of quantum memory or quantum gates in a completely scalable manner. For instance~\cite{Hof05} has discussed estimation methods for the average fidelity based on bounds using classical fidelities on complementary bases. More recently, the use of twirling~\cite{BDSW,DCEL} and randomization methods~\cite{EAZ,ESMR,KLRB,SMKE,MGE} has been shown to provide a scalable method for estimating the eigenvalues of the twirled noisy operation (which includes the average gate fidelity as a special case). It is hoped that other information such as the variance of the fidelity over the twirling/randomizing gate set may provide useful information about the unknown noise model.

In~\cite{KLRB} it is suggested that the variance of the fidelity measured under the proposed randomized benchmarking protocol may provide useful information about the extent to which the noise is coherent (understood here to mean the noise does not consist solely of Pauli errors). While this may be the case for a small number of qubits $n$, we have shown that the variance of the gate fidelity will decrease exponentially quickly in $n$. This implies that an exponentially increasing number of repetitions of the protocol would be required to obtain information about the coherence of the noise, making the method infeasible for even moderately large systems.

Also note that, assuming the noise is effectively independent of the gate set, in order for the variance to be independent of the initial state and the particular choice of randomizing gates, the randomizing gates must comprise a unitary 4-design. Of course using Haar-random gates will produce a variance that depends only on the noise model, however such a protocol is practical for a small number of qubits since implementing Haar-random unitaries is exponentially hard in $n$. Recently, the existence of efficient approximate unitary $4$-designs has been proven~\cite{HL09} and randomizing under such a gate set would provide methods for estimating the variance of the gate fidelity

Even under a single qubit benchmarking protocol that makes use of a gate set which generates a 4-design, our expression for the variance shows that it depends in a non-trivial way on both the diagonal and off-diagonal elements of the $\chi$-matrix. Hence the extent of the coherence of the noise model can not be inferred from an estimate of the variance alone. However, there remains the possibility that the extent of coherence in the noise could be estimated by comparing results from different randomized benchmarking schemes, eg. with and without supplementary Pauli rotations. This would be a worthwhile topic for further investigation.

Another application of our results is in the context of simulating quantum systems on a quantum computer. This is one of the most important potential applications of quantum information processing, and the most likely to be possible in the near term. Of course an important shortcoming of efficient quantum simulation (relative to inefficient simulation on a classical computer) is that not all the information about the simulated system is available upon measurement. This ``readout problem" poses a practical obstacle and raises the question of what, if any, properties of the system may be estimated with a scalable number of repetitions of the simulation.

As a final comment, in the context of studying quantum chaos, it was suggested in~\cite{EWLC} that the characteristics of fidelity decay under perturbation, an important indicator of quantum chaos, could be estimated in an efficient manner. In particular, under the random matrix conjecture for complex and chaotic systems, the fidelity decay can be predicted exactly under any known perturbation, and compared to the observed decay.  An implicit assumption of that argument is that the variance of the fidelity remains small as the system dimension increases so that a reliable estimate of the mean is possible with a scalable number of repetitions.  Our result on bounding the variance shows that this is indeed the case and gives a rigorous justification to that work.

\begin{acknowledgements}

The authors would like to thank David Cory, Iman Marvian and Marcus Silva for many helpful discussions. J.E and E.M acknowledge financial support from NSERC, CIFAR and MITACS.

\end{acknowledgements}

\appendix

\section{Variance for a Single Qubit}\label{sec:Varsingle}

In this section $\text{Var}(\mathcal{F})$ is calculated in a more compact form for the case of a single qubit. Since we already have a simple expression for $\overline{\mathcal{F}}$ given by Eq. (\ref{eq:Fchi}) we only need to calculate $\overline{\mathcal{F}^2}$. We will use Eq. (\ref{eq:Fsquaredtotal}) which will allow us to group particular terms together to obtain a more simple expression.

To begin, we recall some properties of $\chi$. First, $\chi$ is positive semi-definite and has trace equal to 1. Second,
\begin{equation}
\sum_{l,m}\chi_{l,m}P_lP_m = \Lambda(\Id) = d\Lambda\left(\frac{\Id}{d}\right) \nonumber
\end{equation}

\noindent and third,

\begin{equation}\nonumber
\sum_{l,m}\chi_{l,m}P_mP_l =\Id
\end{equation}

\noindent from trace preservation. The 24 terms in Eq. (\ref{eq:Fsquaredtotal}) are sorted into groups of 3 each of which is dealt with separately. Since we are working with a single qubit, $d=2$ in all expressions below. Note that many of the expressions below only hold under the assumption that $d=2$.

\subsection{First Group of Terms}

The first group consists of the following 10 terms:

\begin{gather}
\sum _{l,m,n,r} \chi _{l,m} \chi _{n,r} ([P_l][P_m][P_n][P_r] + [P_l P_r][P_m][P_n] \nonumber \\
+ [P_m P_r][P_l][P_n] +  [P_l][P_m][P_nP_r] + [P_lP_n][P_m][P_r] \nonumber \\ \noalign{\vskip 2mm}
+ [P_lP_n][P_mP_r] + [P_mP_n][P_l][P_r] + [P_mP_n][P_lP_r] \nonumber \\ \noalign{\vskip 2mm}
+ [P_lP_m][P_n][P_r] + [P_lP_m][P_nP_r])\nonumber
\end{gather}

\noindent where for ease of presentation we have used square brackets $``[\: \: \:]"$ to represent the trace operation. Using the assumed properties of the $\{P_i\}$ basis this group can be written as,


\begin{gather}\label{eq:firstgroupofterms}
16\left(\chi _{0,0}^2 + \sum_{l}\chi_{0,l}\chi_{l,0} + \chi_{0,0}\right)+ 8\left(\sum _l\left(\chi_{0,l}^2 + \chi_{l,0}^2\right)\right) \nonumber \\
+ 4\left(\sum_{l,m}\left(\chi_{l,m}^2 + \chi_{l,m}\chi_{m,l} \right) + 1\right).
\end{gather}

\subsection{Second Group of Terms}

The second group consists of the following 8 terms which are grouped as 4 pairs,

\begin{gather}
\sum _{l,m,n,r} \chi _{l,m} \chi _{n,r} (\tr(P_lP_rP_n)\tr(P_m) + \tr(P_lP_nP_r)\tr(P_m) \nonumber \\ \vspace{2mm}
+ \tr(P_mP_rP_n)\tr(P_l)+ \tr(P_m P_nP_r)\tr(P_l) \nonumber\\ \noalign{\vskip 2mm}
+ \tr(P_rP_mP_l)\tr(P_n) + \tr(P_rP_lP_m)\tr(P_n) \nonumber \\ \noalign{\vskip 2mm}
+ \tr(P_nP_mP_l)\tr(P_r) + \tr(P_nP_lP_m)\tr(P_r)).\nonumber
\end{gather}

\noindent The four sums (one for each pair) are calculated independently. For the first sum we deal with five cases:

\bigskip

\noindent Case 1: $n \neq r$, $n\neq0$, and $r\neq0$. This implies $P_nP_r = -P_rP_n$ and so the above is 0.

\bigskip

\noindent Case 2: $n=r$. We get $2\sum_{l,m,n} \chi_{l,m}\chi_{n,n} \tr \left(P_l\right)\tr \left(P_m\right)$ which equals $2\chi_{0,0}d^2$.

\bigskip

\noindent Case 3: $n=0$. We get, $2\sum_{l,m,r} \chi_{l,m}\chi_{0,r}\tr \left(P_lP_r\right)\tr \left(P_m\right)$ which is just $2 \sum_l \chi_{l,0}\chi_{0,l}d^2$.

\bigskip

\noindent Case 4: $r=0$. Similarly to case 3 we get $2 \sum_l \chi_{l,0}^2d^2$.

\bigskip

\noindent Case 5: $r=0$ and $n=0$. This case is required because we have over-counted for this case twice above. The result is $2\chi_{0,0}^2d^2$.

\bigskip

\noindent Hence the five cases give that the first sum is equal to,

\begin{gather}
2\chi_{0,0}d^2 + 2 \sum_l \chi_{l,0}\chi_{0,l}d^2 + 2 \sum_l \chi_{l,0}^2d^2 - 4\chi_{0,0}^2d^2.\nonumber
\end{gather}

The other three sums are calculated in a similar fashion and in total the second group of terms is equal to,

\begin{gather}
8 \chi_{0,0}d^2 + 8 \sum_{l}\chi_{l,0}\chi_{0,l}d^2 + 4 \sum_l\chi_{0,l}^2d^2 \nonumber \\
+ 4 \sum_l\chi_{l,0}^2d^2-16\chi_{0,0}^2d^2.\nonumber
\end{gather}

\noindent Substituting $d=2$ and collecting terms for both the first and second group of terms gives,

\begin{gather}\label{eq:secondgroupofterms}
-48\chi_{0,0}^2 + 48\chi_{0,0}+32\sum_l\chi_{0,l}^2+16\sum_l\chi_{l,0}^2+16\sum_l\chi_{l,0}\chi_{0,l} \nonumber \\
+4\sum_{l,m}\chi_{l,m}^2+4\sum_{l,m}\chi_{l,m}\chi_{m,l} + 4.
\end{gather}

\subsection{Third Group of Terms}

Lastly we have the following 6 terms which are grouped into 3 pairs,

\begin{gather}
\sum _{l,m,n,r} \chi _{l,m} \chi _{n,r}(\tr \left(P_lP_rP_nP_m\right) + \tr \left(P_lP_rP_mP_n\right) \nonumber \\
+ \tr \left(P_lP_nP_mP_r\right)  + \tr \left(P_lP_mP_nP_r\right) \nonumber \\ \noalign{\vskip 2mm}
+  \tr \left(P_lP_mP_rP_n\right) + \tr \left(P_lP_nP_rP_m\right) ). \nonumber
\end{gather}







The first pair is easy to calculate using the same cases as above for m and n. The result is,

\begin{gather}
4\sum_{l,m}\chi_{l,m}\chi_{m,l} + 8\chi_{0,0} - 8\sum_l\chi_{l,0}\chi_{0,l}.\nonumber
\end{gather}

\noindent The second pair requires a bit more effort and we go through the cases separately,

\bigskip

\noindent Case 1: $m\neq n$, $m\neq 0$ and $n\neq 0$. This case gives 0.

\bigskip

\noindent Case 2: $m=n$. In this case the pair becomes $4\sum_{l,m}\chi_{l,m}\chi_{m,l}$.

\bigskip

\noindent Case 3: $m=0$. The pair becomes $2\sum_{l,n,r}\chi_{l,1}\chi_{n,r}\tr \left(P_lP_nP_r\right)$ and after a direct calculation we get,

\begin{gather}
4(\chi_{0,0}  + \chi_{1,0}(\chi_{0,1} + \chi_{1,0} + i\chi_{2,3} -i \chi_{3,2}) \nonumber \\
+ \chi_{2,0}(\chi_{0,2} + \chi_{2,0} - i\chi_{1,3} + i\chi_{3,1}) \nonumber\\
+ \chi_{3,0}(\chi_{0,3} + \chi_{3,0} + i\chi_{1,2} - i\chi_{2,1})).\nonumber
\end{gather}

\medskip

\noindent Case 4: $n=0$. Similar to case 3 we obtain,

\begin{gather}
4(\chi_{0,0}  + \chi_{0,1}(\chi_{0,1} + \chi_{1,0} + i\chi_{2,3} -i \chi_{3,2}) \nonumber \\
+ \chi_{0,2}(\chi_{0,2} + \chi_{2,0} - i\chi_{1,3} + i\chi_{3,1}) \nonumber\\
+ \chi_{0,3}(\chi_{0,3} + \chi_{3,0} + i\chi_{1,2} - i\chi_{2,1})).\nonumber
\end{gather}

\medskip

\noindent Case 5: $m=0$ and $n=0$. This case gives $4\sum_l\chi_{l,0}\chi_{0,l}$.

\bigskip

\noindent Combining the 5 cases gives,

\begin{gather*}
\: \: \: \: \: \: \: \: 4\sum_{l,m}\chi_{l,m}\chi_{m,l} + 8\chi_{0,0} - 8\sum_l\chi_{0,l}\chi_{l,0} \\
\: \: \: \: \: \: + 4\left(\chi_{0,1}+\chi_{1,0}\right)\left(\chi_{0,1}+\chi_{1,0} + i\left(\chi_{2,3}-\chi_{3,2}\right)\right) \\
\: \: \: \: \: \: + 4\left(\chi_{0,2}+\chi_{2,0}\right)\left(\chi_{0,2}+\chi_{2,0} + i\left(\chi_{3,1}-\chi_{1,3}\right)\right) \\
\: \: \: \: \: \: + 4\left(\chi_{0,3}+\chi_{3,0}\right)\left(\chi_{0,3}+\chi_{3,0} + i\left(\chi_{1,2}-\chi_{2,1}\right)\right). \\
\end{gather*}

The third pair can be expressed as,

\begin{gather}
\sum _{l,m,n,r} \chi _{l,m} \chi _{n,r}(\tr \left(P_lP_mP_rP_n\right)  + \tr \left(P_lP_nP_rP_m\right)) \nonumber \\
=\tr \left(\Lambda^{\dagger} \left(\Lambda^{\dagger} \left(\Id\right) \right) \right) + \tr \left(\Lambda\left(\Lambda \left(\Id\right) \right) \right) \nonumber \\
= 4 \nonumber
\end{gather}

\noindent and so combining the three pairs gives,

\begin{gather*}
8\sum_{l,m}\chi_{l,m}\chi_{m,l} + 16\chi_{0,0} - 16\sum_l\chi_{l,0}\chi_{0,l} + 4 \nonumber \\
+ 4\left(\chi_{0,1}+\chi_{1,0}\right)\left(\chi_{0,1}+\chi_{1,0} + i\left(\chi_{2,3}-\chi_{3,2}\right)\right) \nonumber \\
+ 4\left(\chi_{0,2}+\chi_{2,0}\right)\left(\chi_{0,2}+\chi_{2,0} + i\left(\chi_{3,1}-\chi_{1,3}\right)\right) \nonumber \\
+ 4\left(\chi_{0,3}+\chi_{3,0}\right)\left(\chi_{0,3}+\chi_{3,0} + i\left(\chi_{1,2}-\chi_{2,1}\right)\right).
\end{gather*}

We can calculate another expression for the three pairs by noting that four of the terms can be written as,

\begin{gather}
\tr \left(\Lambda \left(\Lambda^{\dagger}\left(\Id\right)\right) \right) + \tr \left(\Lambda^{\dagger} \left(\Lambda\left(\Id\right)\right) \right) \nonumber \\
+ \tr \left(\Lambda \left(\Lambda\left(\Id\right)\right) \right) + \tr \left(\Lambda^{\dagger} \left(\Lambda^{\dagger}\left(\Id\right)\right) \right). \nonumber
\end{gather}

\noindent The above is just $3d + \tr \left(\Lambda^{\dagger} \left(\Lambda\left(\Id\right)\right) \right)$, or even more simply,

\begin{gather}
6 + \tr \left(\Lambda\left(\Id\right)^2 \right).\nonumber
\end{gather}

\noindent The remaining two terms

\begin{equation}
\sum_{l,m,n,r}\chi_{l,m}\chi_{n,r}\tr \left(P_lP_nP_mP_r\right)\nonumber
 \end{equation}

 \noindent and

 \begin{equation}
 \sum_{l,m,n,r}\chi_{l,m}\chi_{n,r}\tr \left(P_lP_rP_mP_n\right)\nonumber
 \end{equation}

 \noindent are complex conjugates of one another. From the calculation of the first pair given above,

\begin{gather}
\sum _{l,m,n,r} \chi _{l,m} \chi _{n,r}(\tr \left(P_lP_rP_nP_m\right) + \tr \left(P_lP_rP_mP_n\right)) \nonumber \\
= 4\sum_{l,m}\chi_{l,m}\chi_{m,l} + 8\chi_{0,0}-8\sum_l\chi_{l,0}\chi_{0,l},\nonumber
\end{gather}

\noindent and since $\sum _{l,m,n,r} \chi _{l,m} \chi _{n,r}\tr \left(P_lP_rP_nP_m\right) = 2$,

\begin{gather}
\sum _{l,m,n,r} \chi _{l,m} \chi _{n,r}\tr \left(P_lP_rP_mP_n\right) \nonumber \\
=4\sum_{l,m}\chi_{l,m}\chi_{m,l} + 8\chi_{0,0}-8\sum_l\chi_{l,0}\chi_{0,l} - 2.\nonumber
\end{gather}

\noindent Therefore the three pairs can also be written as

\begin{gather}
2 + \tr \left(\Lambda\left(\Id\right)^2 \right) + 8\sum_{l,m}\chi_{l,m}\chi_{m,l} + 16\chi_{0,0}-16\sum_l\chi_{l,0}\chi_{0,l}.\label{eq:thirdgroupofterms}
\end{gather}

\noindent Note that by Eq. (\ref{eq:FirstExceptionalTerm}),

\begin{gather}
\tr \left(\Lambda\left(\Id\right)^2\right) = 4\tr\left((\tr_2\chi)^2\right) \nonumber \\
= 2 + 4\left(\chi_{0,1}+\chi_{1,0}\right)\left(\chi_{0,1}+\chi_{1,0} + i\left(\chi_{2,3}-\chi_{3,2}\right)\right) \nonumber \\
\: +  4\left(\chi_{0,2}+\chi_{2,0}\right)\left(\chi_{0,2}+\chi_{2,0} + i\left(\chi_{3,1}-\chi_{1,3}\right)\right) \nonumber \\
\: + 4\left(\chi_{0,3}+\chi_{3,0}\right)\left(\chi_{0,3}+\chi_{3,0} + i\left(\chi_{1,2}-\chi_{2,1}\right)\right).\nonumber
\end{gather}

Combining all 24 terms given in Eq.'s (\ref{eq:firstgroupofterms}), (\ref{eq:secondgroupofterms}) and (\ref{eq:thirdgroupofterms}), and noting $\tr \left(\pi_{\text{sym}}(4,d)\right) = \frac{120}{24}$,

\begin{equation*}
\overline{\mathcal{F}^2} = \frac{
\left(\begin{array}{l}
\ \ -48\chi_{0,0}^2 + 64\chi_{0,0}+24\left(\chi\chi^T + \chi^T\chi\right)_{0,0}\\
+ 32\left(\chi^2\right)_{0,0} + 4\tr \left(\chi\chi^T\right)+12\tr \left(\chi^2\right) \\
+ 6 + 4\tr\left((\tr_2\chi)^2\right).
\end{array}\right)
}{120}
\end{equation*}

\noindent Using the definition of $\chi_0$, and using the expression for the average fidelity given by Eq. (\ref{eq:Fchi}), we have that the variance of the gate fidelity for a single qubit is given by Eq. (\ref{eq:varsinglequbit}).


\section{Variance in Large Dimensions}\label{sec:Large}

To deduce the asymptotic behavior of $\text{Var}\left(\mathcal{F}\right)$ we use the expression for $\overline{\mathcal{F}^2}$ given in Eq. (\ref{eq:Fsquaredchi1}). From this equation one can obtain the following expresssion for $\text{Var}(\mathcal{F})$,

\begin{gather}\label{eq:varchi11}
 \text{Var}\left(\mathcal{F}\right) =  \frac{rd^4+sd^3+ud^2+vd+w}{d\left(d^2+2d+1\right)\left(d^2+5d+1\right)}.
\end{gather}

\noindent where,

\begin{eqnarray*}
r&=&-4\chi_{0,0}^2 + \left(\chi\chi^T\right)_{0,0} + \left(\chi^T\chi\right)_{0,0} + 2\left(\chi^2\right)_{0,0}, \nonumber \\ \noalign{\vskip 2mm}
s&=&-6\chi_{0,0}^2 + \left(\chi\chi^T\right)_{0,0} + \left(\chi^T\chi\right)_{0,0}  + \tr\left(\chi \chi^T\right) - 4\chi_{0,0} \nonumber \\ \noalign{\vskip 2mm}
&\:& + \tr\left(\chi ^2\right) + 2\left(\chi^2\right)_{0,0}, \nonumber \\ \noalign{\vskip 2mm}
u&=&-8\chi_{0,0}+\tr\left(\chi \chi^T\right)+\tr\left(\chi^2\right) \nonumber \\ \noalign{\vskip 2mm}
&\:& +  2\tr\left(\sum_l\left(\chi_{l,0}+\chi_{0,l}\right)P_l\Lambda\left(\Id\right)\right)  -1, \nonumber \\ \noalign{\vskip 2mm}
v&=&2\tr\left(\sum_{l,m}\chi_{l,m}P_l\Lambda(P_m)\right) \nonumber \\ \noalign{\vskip 2mm}
&\:& + 2\tr\left(\sum_l\left(\chi_{l,0}+\chi_{0,l}\right)P_l\Lambda\left(\Id\right)\right) + \tr\left(\left(\Lambda\left(\Id\right)\right)^2\right)  - 3, \nonumber \\ \noalign{\vskip 2mm}
w&=&2\tr\left(\sum_{l,m}\chi_{l,m}P_l\Lambda(P_m)\right) + \tr\left(\left(\Lambda\left(\Id\right)\right)^2\right).
\end{eqnarray*}

\noindent The denominator of (\ref{eq:varchi11}) is a quintic polynomial in $d$. The numerator contains powers of d up to and including $d^4$, however the coefficients depend on $\chi$. We would like to bound these coefficients in terms of $d$.

First, since $\chi$ is a trace-1 positive semi-definite matrix, we obtain the bounds $0 \leq \chi_{0,0}^2 \leq \chi_{0,0} \leq 1$ and $0 \leq \tr\left(\chi^2\right) \leq \tr\left(\chi\right)=1$. Next, for a linear operator A, the Frobenius (Hilbert-Schmidt) norm of A, denoted by $\|\:\:\|_F$, is given by $\|A\|_F=\sqrt{\text{tr}\left(A^{\dagger}A\right)}$. Using the Cauchy-Schwarz inequality we obtain,

\begin{equation}
\left|\tr\left(\chi\chi^T\right)\right|\leq \left\|\chi\right\|_F\left\|\chi^T\right\|_F.\nonumber
\end{equation}

\noindent Since $\chi$ and $\chi^T$ have the same singular values, $\|\chi\|_F=\|\chi^T\|_F$. Therefore $\|\chi\|_F\leq 1 \Rightarrow \left|\tr\left(\chi\chi^T\right)\right|\leq 1$. This also implies $\left|\left(\chi \chi^T\right)_{0,0}\right|\leq 1$ and $\left|\left(\chi^T\chi\right)_{0,0}\right| \leq 1$. To deal with $\Lambda\left(\Id\right)$, we note that it has trace d and is positive semi-definite. Hence $0 \leq \tr\left(\left(\Lambda\left(\Id\right)\right)^2\right) \leq d^2$.

The only two coefficients that remain to be bounded are $\tr\left(\sum_l\left(\chi_{l,0}+\chi_{0,l}\right)P_l\Lambda\left(\Id\right)\right)$ and $\tr\left(\sum_{l,m}\chi_{l,m}P_l\Lambda(P_m)\right)$. By the Cauchy-Schwarz inequality,

\begin{eqnarray}
\left|\tr\left(\sum_l\chi_{l,0}P_l\Lambda\left(\Id\right)\right)\right| &\leq& \left\|\sum_l\chi_{l,0}P_l\right\|_F\left\|\Lambda\left(\Id\right)\right\|_F \nonumber \\
&\leq& d\left\|\sum_l\chi_{l,0}P_l\right\|_F.\nonumber
\end{eqnarray}

\noindent Since,

\begin{eqnarray}
\left\|\sum_l\chi_{l,0}P_l\right\|_F&=&\sqrt{\tr\left(\left(\sum_l\chi_{l,0}P_l\right)^{\dagger}\left(\sum_m\chi_{m,0}P_m\right)\right)} \nonumber \\
&=&\sqrt{\tr\left(\left(\sum_l\chi_{0,l}P_l\right)\left(\sum_m\chi_{m,0}P_m\right)\right)}\nonumber \\
&=& \sqrt{d}\sqrt{\left(\chi^2\right)_{0,0}} \nonumber \\
&\leq& \sqrt{d}\nonumber
\end{eqnarray}

\noindent we get $\left|\tr\left(\sum_l\left(\chi_{l,0}+\chi_{0,l}\right)P_l\Lambda\left(\Id\right)\right)\right| \leq 2d^{\frac{3}{2}}$.

Finally, we need to bound $\tr\left(\sum_{l,m}\chi_{l,m}P_l\Lambda(P_m)\right)$. Using Eq. (\ref{eq:thirdrelate}) we have $\tr\left(\sum_{l,m}\chi_{l,m}P_l\Lambda(P_m)\right)=d^2\tr\left(S \left(S\chi\right)^{T_1} \chi\right)$ where $S$ is the unitary Kraus operator for the SWAP gate. Again, by the Cauchy-Schwarz inequality,

\begin{gather}
\left|\tr\left(\chi S \left(S\chi\right)^{T_1} \right)\right| \leq \left\|\chi S\right\|_F\left\|\left(S\chi\right)^{T_1}\right\|_F \nonumber \\
= \sqrt{\tr\left(\left(\chi S\right)\left(\chi S\right)^{\dagger}\right)}\sqrt{\tr\left(\left(\left(S\chi\right)^{T_1}\right)^{\dagger}\left(S\chi\right)^{T_1}\right)}\nonumber \\
\leq \sqrt{\tr\left(\left(\left(S\chi\right)^{T_1}\right)^{\dagger}\left(S\chi\right)^{T_1}\right)}\nonumber
\end{gather}

\noindent since $\sqrt{\tr\left(\left(\chi S\right)\left(\chi S\right)^{\dagger}\right)}=\sqrt{\tr\left(\chi^2\right)}=\left\|\chi\right\|_F\leq 1$. For any $A$, $B \in L\left(\Hilb\otimes \Hilb\right)$, $\left(A^{\dagger}\right)^{T_1}=\left(A^{T_1}\right)^{\dagger}$ and $\tr\left(\left(AB\right)^{T_1}\right)=\tr\left(B^{T_1}A^{T_1}\right)$. Therefore,

\begin{eqnarray}
\tr\left(\left(\left(S\chi\right)^{T_1}\right)^{\dagger}\left(S\chi\right)^{T_1}\right) &=& \tr\left(\left(\left(S\chi\right)^{\dagger}\right)^{T_1}\left(S\chi\right)^{T_1}\right) \nonumber \\
&=& \tr\left(\left(S\chi\right)^{T_1}\left(\left(S\chi\right)^{\dagger}\right)^{T_1}\right)\nonumber \\
&=& \tr\left(\left(\left(S\chi\right)^{\dagger}\left(S\chi\right)\right)^{T_1}\right) \nonumber \\
&=& \tr\left(\chi^2\right) \leq 1,\nonumber
\end{eqnarray}

\noindent which implies $\left|\tr\left(\sum_{l,m}\chi_{l,m}P_l\Lambda(P_m)\right)\right|\leq d^2$.

Combining all of these results and ignoring negative terms in (\ref{eq:varchi11}) gives,

\begin{eqnarray}
\text{Var}(\cF) &\leq& \frac{|r|d^4+ |s| d^3 + |u|d^2 + |v|d + |w|}{d\left(d^2+2d+1\right)\left(d^2+5d+1\right)}\nonumber \\
&=& \frac{4d^3 + 4d^{\frac{5}{2}} + 9d^2 + 4d^{\frac{3}{2}} + 5d}{\left(d+1\right)^2\left(d+2\right)\left(d+3\right)} \nonumber \\
&\sim& O\left(\frac{1}{d}\right).
\end{eqnarray}

%

%



\end{document}